\documentclass[reprint,
 superscriptaddress,
 amsmath,amssymb,
 aps,
 prb,
]{revtex4-2}

\usepackage{hyperref}
\hypersetup{
    colorlinks=true,
    linkcolor=blue,
    filecolor=blue,      
    urlcolor=blue,
    citecolor=blue
}

\usepackage{cleveref}
\usepackage{bbold}
\usepackage{graphicx}
\usepackage{bm}
\usepackage{physics}

\begin{document}


\title{Photon-mediated interactions and dynamics of coherently driven quantum emitters in complex photonic environments}

\author{A. Miguel-Torcal}
 \email{alberto.miguel@uam.es}
 \affiliation{Departamento de F\'isica Te\'orica de la Materia Condensada, Universidad Aut\'onoma de Madrid, E- 28049 Madrid, Spain.}
 \affiliation{Condensed Matter Physics Center (IFIMAC), Universidad Aut\'onoma de Madrid, E- 28049 Madrid, Spain.}
\author{A. Gonz\'alez-Tudela}
 \affiliation{Institute of Fundamental Physics IFF-CSIC, Calle Serrano 113b, 28006 Madrid, Spain.}%
\author{F. J. Garc\'ia-Vidal}
 \affiliation{Departamento de F\'isica Te\'orica de la Materia Condensada, Universidad Aut\'onoma de Madrid, E- 28049 Madrid, Spain.}
 \affiliation{Condensed Matter Physics Center (IFIMAC), Universidad Aut\'onoma de Madrid, E- 28049 Madrid, Spain.}
\author{A. I. Fern\'andez-Dom\'inguez}
 \affiliation{Departamento de F\'isica Te\'orica de la Materia Condensada, Universidad Aut\'onoma de Madrid, E- 28049 Madrid, Spain.}
 \affiliation{Condensed Matter Physics Center (IFIMAC), Universidad Aut\'onoma de Madrid, E- 28049 Madrid, Spain.}

\date{\today}

\begin{abstract}
In recent years, Born-Markov master equations based on tracing out the electromagnetic degrees of freedom have been extensively employed in the description of quantum optical phenomena originating from photon-mediated interactions in quantum emitter ensembles. The breakdown of these effective models, built upon assumptions such as ensemble spectral homogeneity, an unstructured photonic density of states, and weak light–matter coupling, has also recently attracted considerable attention. Here, we investigate the accuracy of this well-established framework beyond the most conventional, and extensively explored, spontaneous emission configuration. Specifically, we consider a system comprising two coherently driven and detuned quantum emitters, embedded within a hybrid photonic-plasmonic cavity, formed by a metallic nanorod integrated into a high-refractive-index dielectric microresonator. The local density of photonic states in this structure exhibits a complex frequency dependence, making it a compelling platform for exploring photon-mediated interactions beyond the assumptions above. We benchmark this modeling approach for the quantum dynamics of the emitter pair against exact calculations based on a macroscopic field quantization formalism, providing an illustrative assessment of its validity in significantly structured and dispersive photonic environments. Our analysis reveals four distinct regimes of laser driving and frequency splitting that lead to markedly different levels of accuracy in the effective model.   
\end{abstract}

\maketitle


\section{Introduction}

The presence of material structures in the vicinity of an ensemble of quantum emitters (QEs) enables the control over their spontaneous decay~\cite{Quang1994,DKW2000} as well as their mutual photon-mediated interactions~\cite{Agarwal1998}. This environmental influence on the dynamics of QEs is captured by the Born-Markov master equation derived by Düng et al.~\cite{Dung2002}, which is rooted in the framework of macroscopic quantum electrodynamics (QED)\cite{Buhmann2008,Feist2021}. Rather than dealing with the problem of explicitly quantizing the, in principle, unbounded number of optical modes sustained by complex composite structures, this approach consists in tracing out of the electromagnetic (EM) degrees of freedom, thus drastically reducing the dimensionality of the Hilbert space. Under the assumption that the QEs are identical and they are only weakly coupled to their photonic environment, this framework yields a reduced dynamical master equation for the system, with emitter interactions parametrized in terms of the EM dyadic Green’s function~\cite{bookNovotny2012}. In the past few years, owing to its low computational cost and insightful power, this framework has seen widespread use in the description of quantum optical phenomena. Examples of this fruitful research are theoretical proposals for entanglement generation in plasmonic waveguides~\cite{AGT2011,Cano2011} and inverse-designed dielectric cavities~\cite{AMT2022,AMT2024}, the production of distant superradiance in negative-refractive lenses~\cite{Kastel2005} and optical waveguides~\cite{Cardenas2023}, the quantum state manipulation in QE arrays~\cite{Tudela2015,Paz2023,Yelin2024}, the generation of effective magnetic responses~\cite{Rasoul2020} and topological protection~\cite{Elcano2025} in optical metamaterials, the engineering energy transfer in hyperbolic media~\cite{Cortes2017} or the generation of antibunched light in chiral platforms~\cite{Lodahl2017,Downing2019}.     

Recently, photonic devices engineered at the nanoscale have emerged as promising platforms for the miniaturization of quantum technologies, with particular emphasis on their integration with QE ensembles~\cite{Koenderink2015,FJGV2024}. These systems have found applications across diverse domains, including quantum optical circuitry, quantum computation, and quantum sensing~\cite{Lodahl2015,Chang2018}. Among the various nanophotonic architectures explored in this context, metallodielectric structures~\cite{Doeleman2016,Sandoghdar2018} stand out due to their hybrid nature, which enables the combination of the strong subwavelength optical confinement and enhanced light-matter interactions of plasmonic resonances~\cite{Ciraci2012,LiRQ2016} and the long lifetime and large quality factors of Fabry-Perot or microcavity modes~\cite{Pellegrino2020,Albrechtsen2022}. The interplay between the strongly localized, broad resonances and the extended, narrow modes supported by its constituents results in a rich EM environment, featuring asymmetric spectral responses and multiscale spatial field distributions~\cite{Franke2019,Medina2021}.

In this paper, we exploit the near-field EM spectrum provided by a hybrid nanophotonic cavity formed by a noble metal nanorod embedded within a high-refractive-index microsphere to explore the accuracy of the reduced Born-Markov master equation description~\cite{Dung2002} of a pair of QEs coherently driven by a continuous laser. We analyze the range of validity of this modelling scheme with respect to parameters such as the driving strength and the QE-QE and QE-laser detunings. To do so, we benchmark its predictions against exact, computationally more demanding, macroscopic QED solutions for the density matrix of the complete system, accounting for the photonic degrees of freedom through a few-mode quantization approach~\cite{Medina2021,Monica2022}. By incorporating the realistic complexity of the nanophotonic spectral density and a laser driving set-up, we perform a systematic assessment of the reduced master equation model beyond the usual spontaneous emission configuration, whose breakdown and refinement has been the object of much investigation lately~\cite{Schaller2008,Mccauley2020,Fernandez2024}.

The rest of the paper is organized as follows: In Section II, we give a brief overview of the reduced master equation model, along with an exact treatment of macroscopic field quantization. Section III presents the comparative analysis of the results obtained from both theoretical approaches for the quantum nanophotonic system of our choice. Finally, in Section IV, the general conclusions of our study are presented.

\section{Theoretical framework}

We start by briefly introducing the macroscopic QED Hamiltonian that describes exactly the quantum dynamics of two coherently-driven QEs within the electric dipole approximation and in a non-magnetic environment~\cite{Perina2001,Buhmann2008}. It reads (we set $\hbar=1$ in the following)
\begin{multline}
\mathcal{H}=\sum_{j=1}^2\omega_j\sigma_j^\dagger\sigma_j+\int\dd^3r\int_0^\infty\dd\omega\omega\vb{f}^\dagger(\vb{r},\omega)\vb{f}(\vb{r},\omega)-\\
\sum_{j=1}^2\vb{d}_j\vb{E}(\vb{r}_j)+\frac{1}{2}\sum_{j=1}^2[\Omega_j\sigma_j^\dagger e^{-\imath\omega_Lt}+\rm h.c.].
\label{QED_H}
\end{multline}
The QEs are modelled as two-level-systems with annihilation (creation) operators $\sigma_j$ ($\sigma_j^{\dagger}$), which satisfy the relation $[\sigma^{\dagger}_i,\sigma_j]=\delta_{ij}(1-2\sigma^{\dagger}_i\sigma_i)$, where $i,j$ label the QEs. The dipole operators are $\vb{d}_{j}=\vb*{\mu}_j(\sigma_j+\sigma_j^\dagger)$, with transition dipole moments $\vb*{\mu}_j$; and $\vb{r}_j$ and $\omega_j$ are the positions and natural frequencies of the QEs. The electric field operator $\vb{E}(\vb{r})$ is expressed as a linear combination of polaritonic (bosonic) annihilation operators $\vb{f}(\vb{r},\omega)$, according to: $\vb{E}(\vb{r})=\int\dd^3r\int_0^\infty\dd\omega[\vb{G}_e(\vb{r},\vb{r'},\omega)\vb{f}(\vb{r'},\omega)+\rm h.c.]$, with amplitudes determined by $\vb{G}_e(\vb{r},\vb{r'},\omega)=\imath\omega^2/c^2\sqrt{{\rm Im}\{\epsilon(\vb{r},\omega)\}/\pi\epsilon_0}\vb{G}(\vb{r},\vb{r'},\omega)$. Here, $\vb{G}(\vb{r},\vb{r'},\omega)$ is the electric dyadic Green’s function~\cite{bookNovotny2012} for the photonic environment of the QEs, which captures the full spatial and frequency-dependent response of the surrounding medium, characterized by the permittivity $\epsilon(\vb{r},\omega)$. The last term in Eq.~\eqref{QED_H} accounts for the laser driving on each QE, with frequency $\omega_L$, and whose amplitude is characterized by $\Omega_j$~\cite{Cano2011,bookHaroche2006}, which we refer to as coherent driving amplitudes.

The Hamiltonian above involves a 4-dimensional continua of polaritonic operators, labeled by $\{\vb{r},\omega\}$. The intractably large Hilbert space associated makes it unfeasible to directly apply Eq.~\eqref{QED_H} for calculations involving complex quantum nanophotonic systems. Circumventing this limitation, Düng \emph{et al.} demonstrated that, in the case of an ensemble of QEs with identical natural frequencies ($\omega_i=\omega_0$ for all $i$) which are only weakly coupled to their dielectric environment, the Heisenberg equation for the Hamiltonian in Eq.~\eqref{QED_H} could be casted, after tracing out the photons from the system, into a Born-Markov master equation for the reduced density matrix describing the quantum state of the QEs~\cite{Dung2002} 
\begin{equation}
\dot{\rho}=\imath[\rho,H]+\sum_{i,j=1}^2\frac{\gamma_{ij}}{2}\mathcal{L}_{\sigma^{\dagger}_i,\sigma_j}[\rho],
\label{MEq}
\end{equation}
where $\mathcal{L}_{O,O'}=2O'\rho O-OO'\rho-\rho OO'$ are standard Lindblad superoperators. Importantly, the master equation above inherently assumes that the photon bath in contact with the QE ensemble has an unstructured energy spectrum~\cite{bookPetruccione2007}. For laser-driven QEs, which is the configuration of interest, the Hamiltonian $H$ in Eq.~\eqref{MEq} can be written in the rotating frame of the laser as~\cite{AMT2024,AVV2021}
\begin{equation}
H=\sum_{i=1}^2\delta_i\sigma_i^\dagger\sigma_i+\sum_{i\neq j=1}^2
g_{ij}\sigma_i^\dagger\sigma_j+\sum_{i=1}^2\Omega_i(\sigma_i+\sigma_i^\dagger),
\label{MEq_H}
\end{equation}
where $\delta_i=\omega_i-\omega_{\rm L}$ is the detuning of the QE's natural frequency with respect to $\omega_{\rm L}$. Note that here we allow the natural frequencies of the QEs to be different, as we will explore the effect of emitter detuning in our study. Eqs.~\eqref{MEq} and~\eqref{MEq_H} show that emitter-emitter photon-mediated interactions are separated into two different contributions: the dissipative coupling assisted by on-resonant EM fields, with frequency matching the QEs frequency (assumed to be identical in Ref.~\cite{Dung2002}); and the coherent coupling assisted by the bath of off-resonant EM fields. These are weighted by the interaction strengths $\gamma_{ij}$ and $g_{ij}$, respectively, whose expressions in terms of the Dyadic Green's function, evaluated at the QE position and natural frequency, are
\begin{eqnarray}
\gamma_{ij}=\frac{2\omega_0^2}{\epsilon_0c^2}\vb*{\mu}_i{\rm Im}\{\vb{G}(\vb{r}_i,\vb{r}_j,\omega_0)\}\vb*{\mu}_j, \label{gamma} \\
g_{ij}=\frac{\omega_0^2}{\epsilon_0c^2}\vb*{\mu}_i{\rm Re}\{\vb{G}(\vb{r}_i,\vb{r}_j,\omega_0)\}\vb*{\mu}_j.
\label{g}
\end{eqnarray}
The term weighted by $\gamma_{ii}$ in Eq.~\eqref{MEq} simply describes the spontaneous radiative decay of each emitter. In their original work, D\"ung \emph{et al.} briefly noted that their framework could be extended to QEs with different natural frequencies by setting $\omega_0=(\omega_i+\omega_j)/2$ in Eqs.~\eqref{gamma} and \eqref{g}. While the substitution is formally straightforward, it implicitly assumes that the frequency difference $\abs{\omega_i-\omega_j}$ is small compared to the characteristic frequency scale over which the dyadic Green’s function varies significantly. Here, we will examine the validity of this approximation.

The appearance of sharp resonances in the Green's function, as well as the presence of QEs with distant transition frequencies and the effect of their Rabi dressing by the driving laser, can lead to the breakdown of the assumption above. The purpose of our work is to analyze in depth the range of validity of Eq.~\eqref{MEq}. To do so, we benchmark its predictions with those obtained from the exact solution of Eq.~\eqref{QED_H}, which is constructed using a few-mode quantization scheme that we recently developed, see Refs.~\cite{Medina2021} and \cite{Monica2022} for a detailed derivation. This scheme casts the quantum dynamical evolution for the system into a master equation for the density matrix accounting for both the QE and photonic degrees of freedom, the latter restricted to a reduced number of coherently interacting optical modes, each coupled to the QEs and to an independent Markovian bath. 
This approach is able to produce an exact solution for the original problem~\cite{HuelgaPlenio2018}, but requires an accurate parametrization of the master equation through the fitting of the tensorial spectral density, $\vb{J}(\omega)$, whose entries are given by
\begin{equation}
J_{ij}(\omega)=\frac{\omega^2}{\pi\epsilon_0c^2}\vb*{\mu}_i{\rm Im}\{\vb{G}(\vb{r}_i,\vb{r}_j,\omega)\}\vb*{\mu}_j, \\
\label{J-spectral}
\end{equation}
where $i$ and $j$ label the QEs. This function encodes the light-matter interactions taking place in the system~\cite{bookPetruccione2007}, and recover $J_{ij}(\omega_0)=\gamma_{ij}/2\pi$ in Eq.~\eqref{gamma}.

\section{Results}
\begin{figure}
\centering
\includegraphics[width=\linewidth]{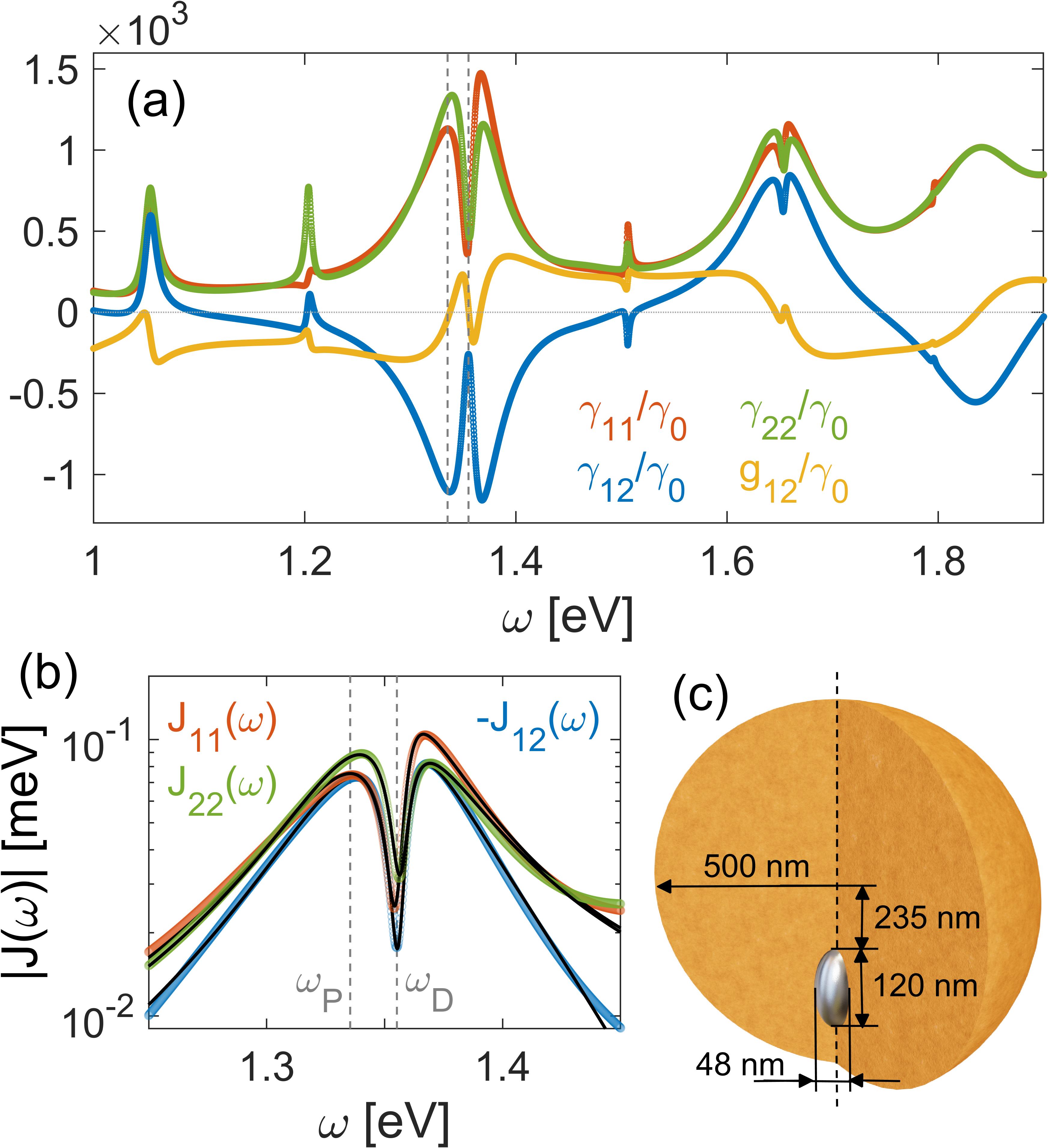}
\caption{(a) Dissipative and coherent coupling parameters and QE decay rates as a function of frequency (1-2 eV), normalized to the QE free-space decay rate. (b) Color dots plot the diagonal and non-diagonal entries of the generalized spectral density, $\vb{J}(\omega)$. Black lines correspond to their 3-interacting-modes fitting for $\omega\in[1.25,1.45]$ eV. Vertical dashed lines indicate the peak (P) and dip (D) frequencies. (c) Schematic of the hybrid metallodielectric structure and its key geometrical parameters. The emitters are located 10 nm above (QE1) and below (QE2) the nanorod, and oriented along the radial direction.}
\label{fig:1}
\end{figure}

To explore photon-mediated interactions between a pair of QEs and their influence on population dynamics, we consider a complex yet feasible hybrid metallodielectric platform comprising a plasmonic nanocavity and a high-refractive-index microresonator, sketched in Figure~\ref{fig:1}(c). The set-up is composed of a $500$ nm radius GaP sphere ($\varepsilon=9$~\cite{Maier2017}) supporting multiple long-lived and delocalized Mie resonances in the frequency range of interest (1-2 eV).  A 120 nm long, 48 nm wide silver nanorod, with permittivity taken from Ref.~\cite{AFD2013}, is embeded within it, located at a position displaced 235 nm from the sphere center. This metallic element sustains a few, tightly confined localized surface plasmons in the same frequency window. The QE positions are chosen to ensure the probing of the hybrid spectrum of the composed structure, 10 nm above (QE1) and below (QE2) the nanorod, with their dipole moments parallel to the radial direction. Their moduli are set within a realistic range, $\mu_1=\mu_2=0.83~e\cdot{\rm nm}$, corresponding to single organic molecules or quantum dots~\cite{Lester2000}.

As discussed above, the photon-mediated interaction between the two QEs, given by the dissipative and coherent coupling strengths in Eqs.~\eqref{gamma} and \eqref{g}, are obtained from classical EM calculations for the Dyadic the Green's function, performed with the Maxwell equation solver implemented in the software COMSOL Multiphysics$^{\rm TM}$. Figure~\ref{fig:1}(a) plots the coupling parameters (dissipative in blue, coherent in yellow) and spontaneous decay rates (orange and green) versus frequency (here we make $\omega_0=\omega$), normalized to the emitters decay rate in free space, $\gamma_0=\frac{\omega^3|\vb*{\mu}|^2}{3\pi\hbar\varepsilon_0 c^3}$. The latter correspond to the Purcell enhancement factors experienced by both emitters, reaching values approaching $1.5\cdot10^3$ around 1.35 eV. These master equation parameters reveal the rich photonic spectrum of the specific metallodielectric structure across a broad frequency range, combining sharp photon-like and broad plasmon-like resonances, along with Fano-like features arising from their interaction. Figure~\ref{fig:1}(b) renders the diagonal and non-diagonal entries of the spectral density within the narrow frequency window between 1.25 and 1.45 eV, where the Purcell factors in panel (a) are the largest, which is the region we are most interested in. $J_{ij}(\omega)$ are plotted in absolute value and in semilogarithmic scale, with colors matching those of the related coupling parameters in Figure~\ref{fig:1}(a). The black lines correspond to the fit to $\vb{J}(\omega)$ using three lossy, interacting bosonic modes. The explicit form of the fitting function is provided in the Supplemental Material of Ref.~\cite{Medina2021}. This fit serves as the basis for constructing the full density matrix of the system. We can observe that, despite the reduced number of modes considered, the fitting reproduces faithfully the spectral density, with small deviations only at $\omega\sim 1.45$ eV. In Figure~\ref{fig:1}(a) and (b), the vertical dashed lines indicate the position of the spectral peak and dip, $\omega_{\rm P}=1.335$ eV and $\omega_{\rm D}=1.355$ eV.

\begin{figure}
\centering
\includegraphics[width=\linewidth]{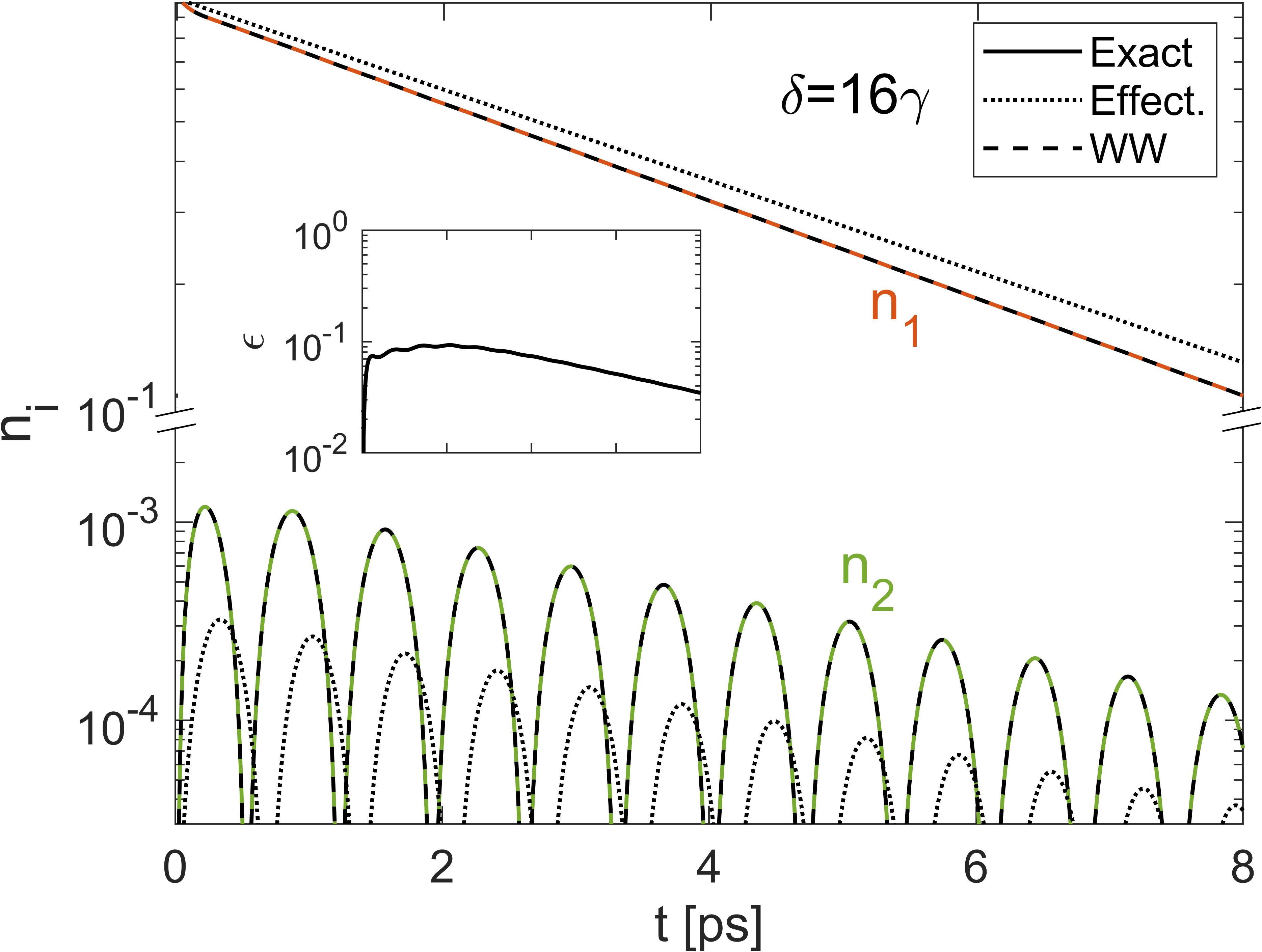}
\caption{Excited-state population dynamics for the spontaneous emission of QE1 (on top of the gold nanorod) with QE2 initially in the ground state. The QE frequencies are set to $\omega_1=\omega_{\rm D}+\delta$ and $\omega_2=\omega_{\rm D}-\delta$. Color solid lines plot the exact solution, obtained through the 3-mode fitting of the spectral density in Figure~\ref{fig:1}, while black dotted lines render the effective solution for the reduced density matrix given by Eq.~\eqref{MEq}. To benchmark both approaches, the Wigner-Weisskopf (WW) calculation for the exact spectral density tensor is shown in black dashed lines. The inset renders the error measure, $\epsilon$, given by Eq.~\eqref{frob} as a function of time.}
\label{fig:2}
\end{figure}

Once we have characterized the nanophotonic platform for our study, we investigate the quantum dynamics of the QE excited-state populations in a spontaneous emission configuration, this is, in the absence of external driving ($\Omega_i=0$ in Eqs.~\eqref{QED_H} and \eqref{MEq_H}). This scenario allows us to test the accuracy of our exact solution based on the three interacting-mode quantization. Figure~\ref{fig:2} renders the population of both QEs versus time when the QE1 is initially excited, while QE2 is in the ground state, $n_1(0)=1$ and $n_2(0)=0$. The QE frequencies are chosen around $\omega_0=\omega_{\rm D}$, at the dip of the Fano-like resonance in Figure~\ref{fig:1} to ensure sufficient variation in the spectral density and thereby test the models in the most general possible scenario. Likewise, the spectral distance between them is chosen to $\omega_2-\omega_1=2\delta$, with $\delta=16\gamma$, where $\gamma=\sqrt{\gamma_{11}\gamma_{22}}=0.48$ meV is the collective emission rate of the QE pair~\cite{Ficek2002}. Three different calculations are compared in Figure~\ref{fig:2}. First, and as reference, black dashed lines plot the Wigner-Weisskopf~\cite{bookPetruccione2007} solution for the exact $\vb{J}(\omega)$, extended to two QEs in a similar way as in Ref.~\cite{Cuartero2018}. Second, $n_{1,2}(t)$ obtained from the three-mode quantization model parametrized with the fitted spectral density in Figure~\ref{fig:1}(b) are rendered in color solid lines (calculated using QuTiP~\cite{QuTip1,QuTiP2}). Finally, the effective solution for the QE populations calculated from Eq.~\eqref{MEq} (also using QuTip) is shown in black dotted lines. The perfect agreement between the first two sets of results, even for $n_2(t)$, allows us to consider the three-mode solution as an exact description of the system (for the QE parameters considered). On the contrary, the effective solution leads to population dynamics that present slight deviations from the exact one, despite the fact that $n_1(t)$ shows an exponential decay characteristic of the weak-coupling regime. These originate from the $2\delta$ detuning between the QEs, and reveals that the approximation $\omega_0=\omega_1=\omega_2\simeq\omega_{\rm D}$ inherent to Eqs.~\eqref{gamma} and \eqref{g} fails. 

In order to systematically compare the predictions of Eq.~\eqref{MEq} with the corresponding exact solutions for the reduced density matrix including the quantized modes, we employ an error measure, given by the difference between the calculated density matrix and the exact one, and defined in terms of the Frobenius norm as  
\begin{equation}
\epsilon=\norm{\rho-\rho^{\rm exact}}_F=\sqrt{\sum_{i,j=1}^2\abs{\rho_{ij}-\rho^{\rm exact}_{ij}}^2},    
\label{frob}
\end{equation}
which is invariant under unitary operations and bounded, given that $\norm{\rho}_F=\sqrt{{\rm Tr}(\rho^2)}\leq 1$. The inset of Figure~\ref{fig:2} displays this magnitude for the spontaneous emission configuration discussed above. Notably, $\epsilon$ remains above 0.05 throughout the whole evolution, showing that the QE detuning, $2\delta/\omega_{\rm D}\simeq0.006$, lead to deviations in the reduced density matrix of the order of $5-10\%$ at all times.

Having tested our theoretical tools in a spontaneous emission configuration, we next consider the case of coherent driving, where both QEs are excited by an identical external laser field, i.e., $\Omega_1=\Omega_2=\Omega$ in Eqs.~\eqref{QED_H} and \eqref{MEq_H}. These driving amplitudes are proportional to the dipole moment of the QE transition and the laser field amplitude at the QE position. Another driving parameter comes into play in the description of the quantum dynamics of the system, the laser frequency, which we restrict to two different values: at the lowest frequency peak in the spectral density, $\omega_{\rm L}=\omega_{\rm P}$, and at the Fano dip next to it, $\omega_{\rm L}=\omega_{\rm D}$ (both indicated as gray vertical lines in Figure~\ref{fig:1}). Our objective is to analyze the evolution of the quantum state of the QE pair under laser driving of increasing strength and detuning. 
\begin{figure}
\centering
\includegraphics[width=\linewidth]{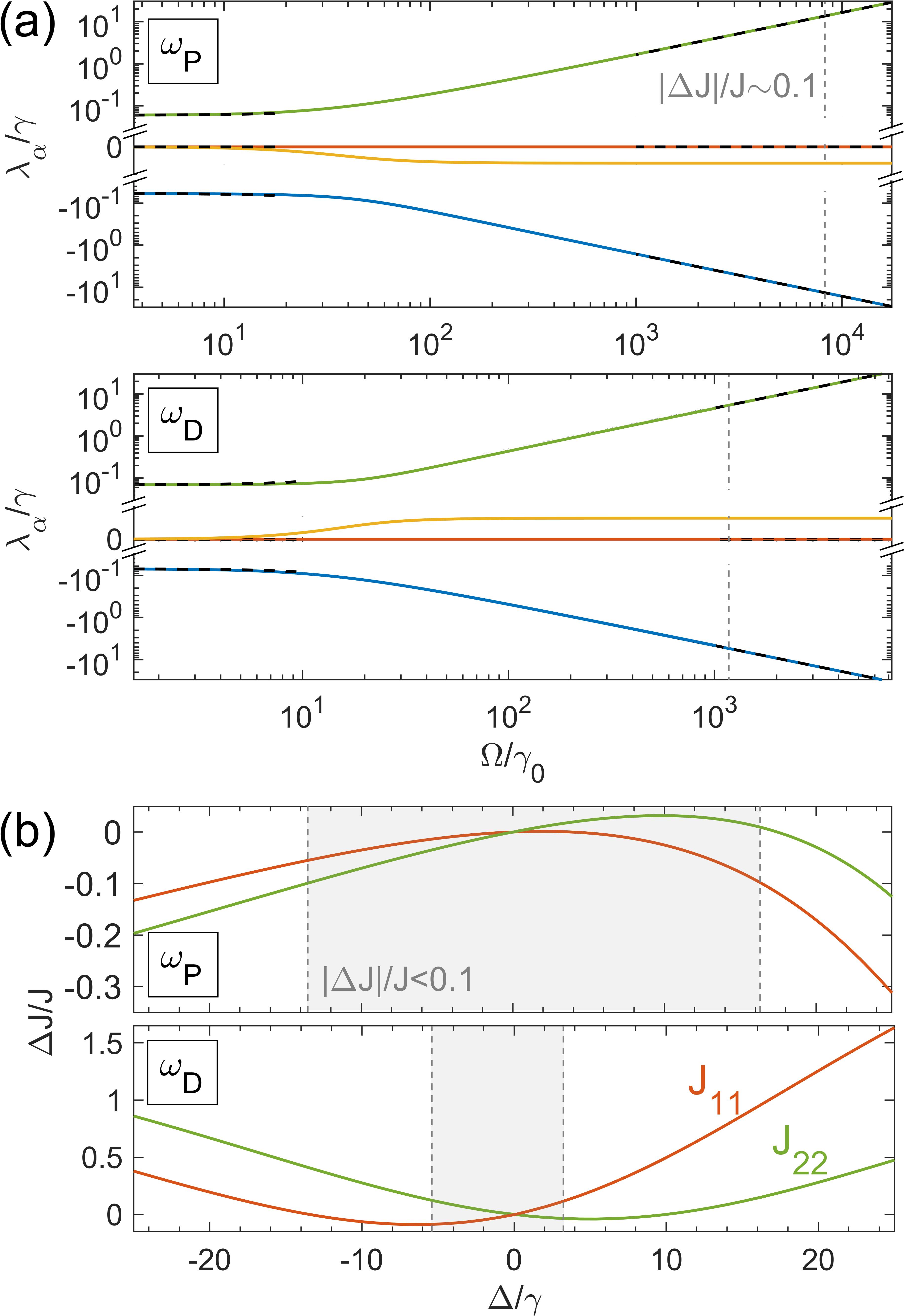}
\caption{(a) Eigenvalues of the Hamiltonian in Eq.~\eqref{MEq_H} (normalized to $\gamma$) as a function of the driving strength (colors). Dashed lines plot the asymptotic analytical expressions discussed in the text. Two different values of the laser frequency, $\omega_{\rm P}$ and $\omega_{\rm D}$ are considered. (b) Relative variation of the spectral density versus generalized Rabi splitting (normalized to $\gamma$) for the laser frequencies in (a). The gray vertical lines (a) and grey shaded areas (b) indicate an estimated threshold of validity of Eq.~\eqref{MEq}, which we tentatively set at $|\Delta J|/J=0.1$.}
\label{fig:3}
\end{figure}

Before proceeding with our numerical analysis, we gain some physical insight into the influence of the laser driving in the system dynamics. From the Hamiltonian in Eq.~\ref{MEq_H}, the energies of the new, laser-dressed eigenstates, are readily obtained as the roots of the characteristic polynomial $\lambda(\lambda^3-\lambda(\Delta^2+4\Omega^2)+4\Omega^2g_{12})=0$, where $\Delta^2=\delta^2+g_{12}^2$~\cite{AVV2021}. In the limit of low driving ($\Omega\rightarrow0$), the dressed eigenenergies tend to those of the bare system, $\lambda=\{0,\pm\Delta\}$, while in the limit of large driving and vanishing coherent interactions ($\Omega\gg\Delta$, $\Omega^2g_{12}\rightarrow 0$), they become $\lambda=\{0,\pm R_\Omega\}$, where $R_\Omega=\sqrt{\Delta^2+4\Omega^2}$ is the generalized Rabi splitting induced by the laser. Color lines in Figure~\ref{fig:3}(a) plot in log-log scale the eigenenergies $\lambda_\alpha$ (obtained numerically and normalized to $\gamma$) as a function of the driving strength (in units of $\gamma_0$). The asymptotic trends extracted analytically are also shown in black dashed lines, reproducing the numerical dispersion except for the yellow curve at high driving, where the discrepancy originates from the finite value of $4\Omega^2g_{12}$. The top and bottom panels correspond to laser frequencies resonant with $\omega_{\rm P}$ and $\omega_{\rm D}$, respectively. The QEs are detuned by $\pm\delta$, as defined above, placing the laser frequency exactly midway between their transition frequencies. They reveal the extent of the energy shifts induced by the driving field, which leads to a significantly larger splitting between the relevant system energies compared to the spontaneous emission configuration. Thus, the QE pair explores the complex spectral density of the hybrid metallodielectric structure in a broader frequency window than in the spontaneous emission case.

The vertical dashed lines in Figure~\ref{fig:3}(a) mark the generalized Rabi splitting for which the relative variation of the spectral density, defined as $|\Delta J|/J={\rm max}\{|J_{jj}(\omega\pm R_\Omega)-J_{jj}(\omega)|\}/J_{jj}(\omega)$, is $10\%$. Note that the value of $R_\Omega\simeq|\lambda_\alpha^{\rm max}-\lambda_\alpha^{\rm min}|/2$ giving rise to this variation is significantly smaller for $\omega_{\rm L}=\omega_{\rm D}$, as $\vb{J}(\omega)$ presents a much sharper dependence on frequency in this case. Therefore, we can expect that Eq.~\eqref{MEq} fails beyond this regime, as it inherently assumes a constant spectral density within the frequency window explored by the QEs. To asses more accurately $|\Delta J|/J$, this magnitude is plotted in Figure~\ref{fig:3}(b) as a function of the generalized splitting $R_\Omega/\gamma$. Now, $J_{11}(\omega)$ (orange) and $J_{22}(\omega)$ (green) are analyzed separately. Again, two laser frequencies, $\omega_{\rm P}$ (upper panel) and $\omega_{\rm D}$ (lower panel) are considered. We can observe that, indeed, the variation of the spectral density in the vicinity of $\omega_{\rm P}$ is rather less pronounced than at $\omega_{\rm D}$. These panels allow us to tentatively estimate $\abs{\Delta J}/J\sim 0.1$ (shaded area) as a region of generalized Rabi splitting within which the tracing out of the photonic degrees of freedom can yield a fairly accurate description of QE-QE interactions in the system.
\begin{figure*}
\centering
\includegraphics[width=.49\linewidth]{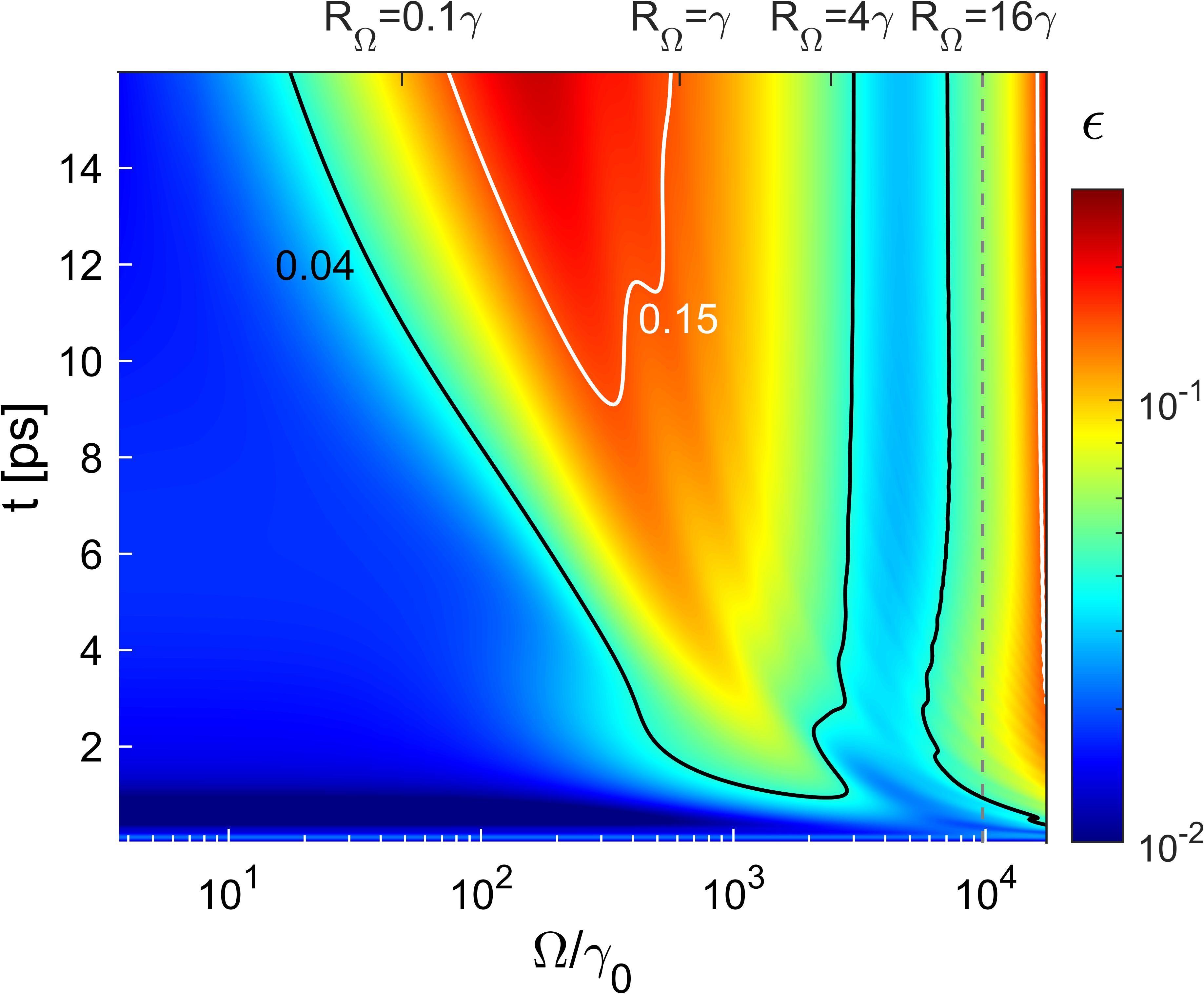}
\includegraphics[width=.49\linewidth]{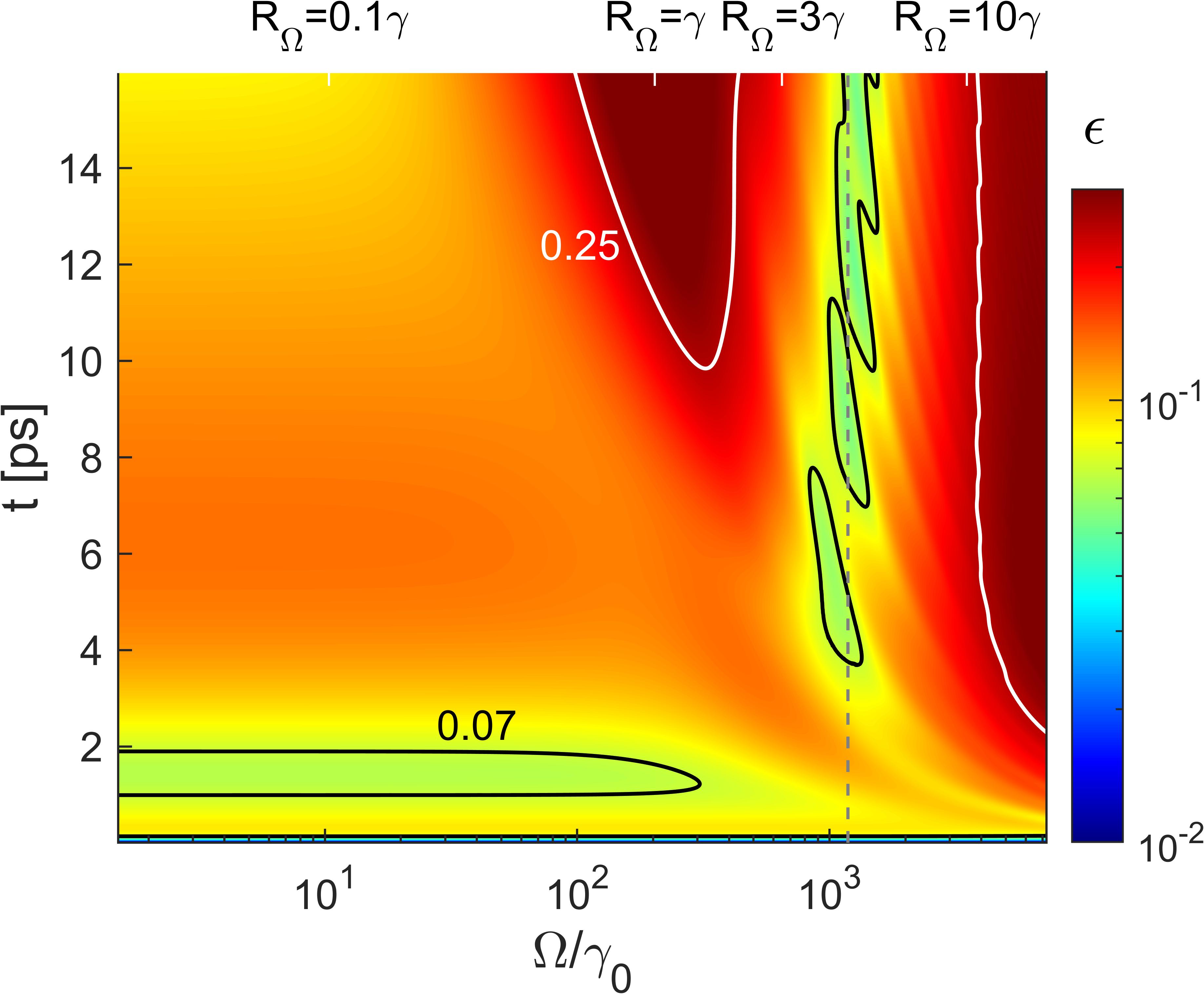}
\caption{Error measure, $\epsilon$, as a function of time and coherent driving amplitude (in units of the emission rate in free space $\gamma_0$) for two different laser frequencies, $\omega_L=\omega_P$ (left panel) and $\omega_L=\omega_D$ (right panel). It quantifies the relative distance between density matrices resulted from applying exact and effective models when solving the excited-state population dynamics in our hybrid structure. Particular relevant values of the generalized Rabi splitting $R_\Omega$ are indicated in the top axis. Black and white solid lines correspond to time-driving configurations yielding two different values of the error measure in each panel.}
\label{fig:4}
\end{figure*}

We have shown that the generalized Rabi splitting, $R_\Omega$, and the relative variation of the spectral density, $|\Delta J|/J$, can be useful ingredients to understand the influence of the laser driving in the emitter-emitter interactions and the temporal evolution of the quantum state of the QE pair. In the following, we numerically solve the effective and exact quantum dynamical equations and compare the reduced density matrices over time through the error measure, $\epsilon$, defined in Eq.~\eqref{frob}. The color maps in Figure~\ref{fig:4} display $\epsilon$ versus time and driving strength (normalized to $\gamma_0$ and in logarithmic scale) with QE1 initially in its excited state, the same initial state as in the spontaneous emission configuration in Figure~\ref{fig:2}. The left (right) panel corresponds to laser frequency at $\omega_{\rm P}$ ($\omega_{\rm D}$) and QE frequencies $\omega_1=\omega_{\rm L}+\delta$ and $\omega_2=\omega_{\rm L}-\delta$. For clarity, the color scale encoding the error measure is saturated in both panels, ranging only from $0.01$ to $0.3$. Thus, dark blue (red) regions indicate the best (poorest) agreement between the effective and exact density matrices. Black and white solid lines denote isocurves of constant $\epsilon$, with specific values reflected in each case. Note that in the upper axes, different values of the generalized Rabi splitting, $R_\Omega$, are indicated, and the vertical, grey, dashed lines mark the condition $|\Delta J|/J=0.1$.

At first glance, Figure~\ref{fig:4} reveals a significantly larger error measure, up to one order of magnitude on average, for $\omega_{\rm L}=\omega_{\rm D}$. This highlights the strong sensitivity of the system dynamics to the EM environment and underscores the limitations of Eq.~\eqref{MEq} to reproduce $\rho(t)$ accurately when the spectral density is strongly structured. We can also identify four distinct regimes for $\epsilon$ depending on the coherent driving strength $\Omega/\gamma_0$, which determines the spectral separation between the optical transitions in the QE pair (these are specially apparent in the left panel). A degenerate regime, in which the generalized Rabi splitting is negligible ($R_\Omega/\gamma \rightarrow 0$); a near-degenerate regime, where it is comparable to the collective emission rate ($R_\Omega \sim \gamma$); a non-degenerate regime, where $R_\Omega > \gamma$ but the relative variation of the spectral density is small, $\abs{\Delta J}/J\lesssim 0.1$; and finally, a far-detuned regime, where $R_\Omega>16\gamma$ and $\abs{\Delta J}/J>0.1$. The effective master equation provides an accurate description of the system in both the degenerate and non-degenerate regimes, where $\epsilon < 0.1$ and $\epsilon < 0.2$ in the left and right panels, respectively. Conversely, the approach fails in the near-degenerate and far-detuned regimes.

In particular, in the degenerate regime, the average frequency approximation, $\omega_0\simeq(\omega_1+\omega_2)/2$, performs best, since the QE transition frequencies are sufficiently close and the spectral density remains largely constant. As the driving strength increases, the error measure grows smoothly, entering the near-degenerate regime. In this regime, the approximations made in Eq.~\eqref{MEq}---namely, the assumption of a flat photonic bath spectrum inherent to the Born-Markov treatment of the effective model~\cite{Mccauley2020}---and the evaluation of Eqs.~\eqref{gamma} and~\eqref{g} at the average frequency fail to capture additional optical transitions between closely spaced energy eigenstates of the dressed QE pair~\cite{Fernandez2023}, thereby leading to a misestimation of the interplay between coherences and populations during the time evolution of near-degenerate states. The non-degenerate regime emerges more abruptly, when the generalized Rabi splitting is large enough to inhibit those transitions, and $\epsilon$ decreases within a narrow range of laser strengths. Finally, for even larger driving, Eq.~\eqref{MEq} overestimates the interaction between the QEs, which now experience well-separated regions of the spectral density $\vb{J}(\omega)$, leading to a rapid increase in the error measure.

\begin{figure}[t!]
\centering
\includegraphics[width=\linewidth]{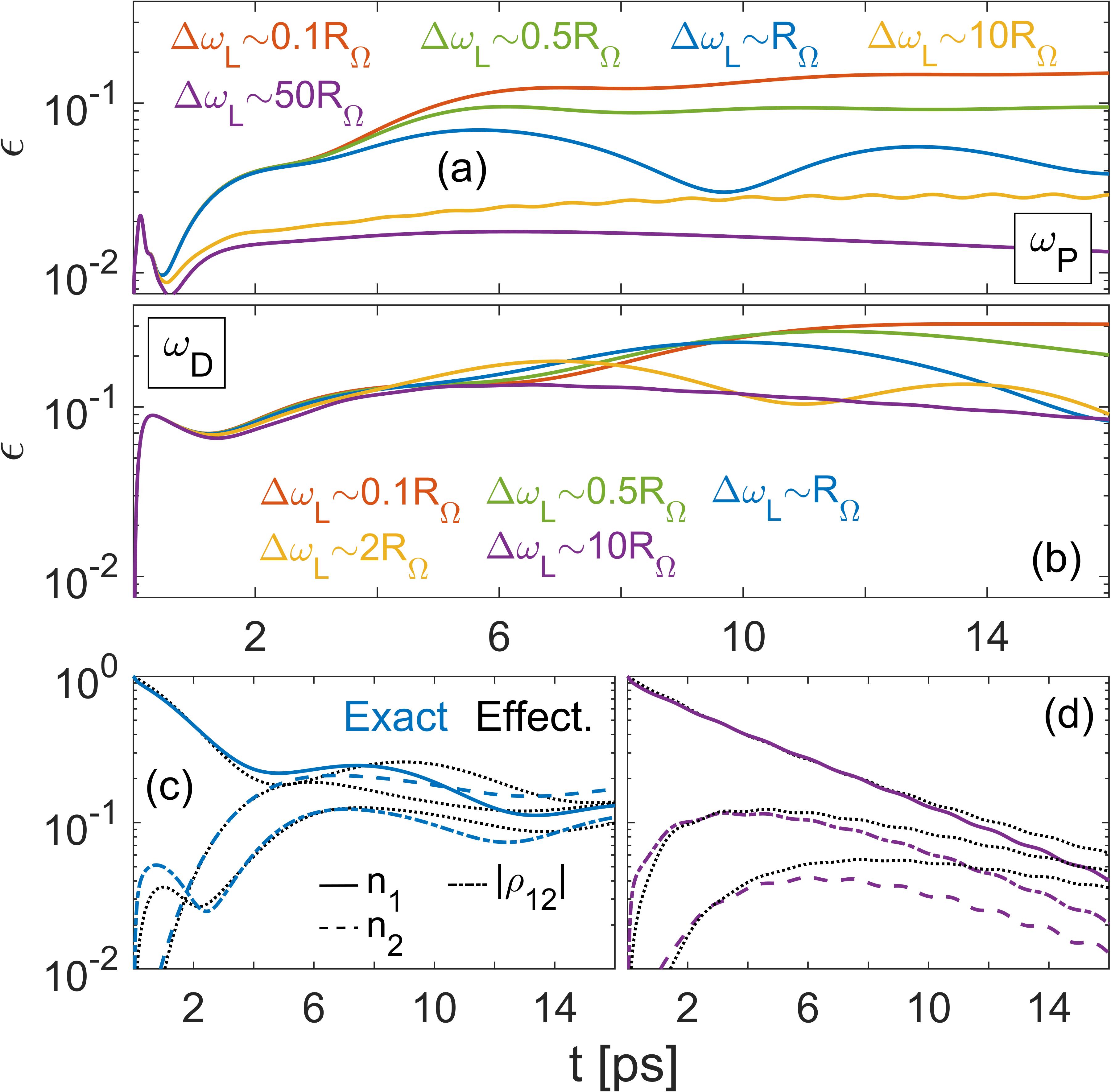}
\caption{(a) Error measure versus time (in color solid lines) for $\omega_{1,2}=\omega_{\rm P}\pm\delta$, $\Omega/\gamma_0=500$, and laser detunings, $\Delta\omega_{\rm L}=\omega_{\rm L}-\omega_{\rm P}$, between 0.1 and $50 R_\Omega$. (b) Same as (a) but for $\omega_{1,2}=\omega_{\rm D}\pm\delta$, $\Omega/\gamma_0=300$ and $\Delta\omega_{\rm L}=\omega_{\rm L}-\omega_{\rm D}$ between 0.1 and $10 R_\Omega$. (c) and (d) plot the QE populations as well as the main coherence in the reduced density matrix as a function of time, calculated with the exact (color lines) and effective (black dotted lines) approaches. Both panels correspond to system configurations for laser in the vicinity of the spectral density dip, and blue (c) and violet (d) colors indicate the value of $\Delta\omega_{\rm L}$ following the same color code as in panel (b).}
\label{fig:5}
\end{figure}

Up to here, we have examined the error measure for increasing coherent driving strength and two different driving frequencies, with the natural frequencies of the QEs symmetrically detuned from the laser. We now proceed in a different way: we fix the laser amplitude, $\Omega$, and evaluate the error measure as a function of the QEs-laser detuning.  Figure~\ref{fig:5} presents the results of our study. The QE frequencies are set to $\omega_{\rm P}\pm\delta$ in panel (a), and $\omega_{\rm D}\pm\delta$ in panel (b), but the laser frequency is no longer at resonance with the peak/dip of the spectral density, but detuned from it by $\Delta\omega_{\rm L}=\omega_{\rm L}-\omega_{\rm P/D}$. The color lines plot the time evolution of $\epsilon$ for five different values of the laser detuning (expressed in units of $R_\Omega$). The laser amplitudes for both cases under study, $\omega_{\rm P/D}$, are chosen to lie within the near-degenerate regime described above ($R_\Omega\sim\gamma$, $\Delta\omega_{\rm L}=0$): $\Omega/\gamma_0=500$ (a) and $\Omega/\gamma_0=300$ (b). The strong dependence of the quantum dynamics on the EM environment of the QEs encoded in $\vb{J}(\omega)$ is also evidenced in these two panels. Note that similarly to Figure~\ref{fig:4}, we find a difference of roughly one order of magnitude on the time-averaged $\epsilon$ between them. 

Figure~\ref{fig:5}(a) and (b) show that as $\Delta\omega_{\rm L}$ increases, the error measure decreases, a trend that is more apparent at longer times, which can be related to the imbalance in the absolute value of the detunings $|\omega_1-\omega_ {\rm L}|$ and $|\omega_2-\omega_ {\rm L}|$. This results in one of the QEs being less efficiently excited, causing the system’s dynamics to be predominantly governed by the other QE, which in turn makes emitter-emitter interactions less relevant. Thus, based on the discussion above, we infer that the laser detuning effectively drives the system out of the near-degenerate regime and into the non-degenerate one. In addition to this gradual decrease of the error measure with increasing laser detuning, we can observe a reduction in the amplitude and period of the oscillations in $\epsilon$ in the transient regime before it reaches a flat plateau at longer times. This trend is similar in both panels. For this reason, we analyze it only for the data in Figure~\ref{fig:5}(b). Figures~\ref{fig:5}(c) and (d) plot the population of the QEs ($n_1$ and $n_2$), as well as the absolute value of the coherence in the first excitation manifold, $|\rho_{12}(t)|$, as a function of time. The colors, blue in panel (c) and purple in panel (d) indicate the value of $\Delta\omega_{\rm L}$, using the same color code as in panel (b). We can observe that the strong and fast oscillating character of the error measure at $\Delta\omega_{\rm L}\sim R_\Omega$ is directly inherited from the Rabi oscillations experienced by the entries of the reduced density matrix, which originate from the emitter-emitter interactions taking place in the system. On the contrary, QE populations decay monotonically at $\Delta\omega_{\rm L}\sim 10 R_\Omega$, which translates into a smoother dependence of $\epsilon$ on time, which also improves the accuracy of the effective description given by Eq.~\eqref{MEq}.

\section{Conclusions}
In this work, we have systematically assessed the validity of the reduced Born-Markov master equation model for the dynamics of a coherently driven quantum emitter pair embedded in complex nanophotonic environment. By considering a hybrid photonic-plasmonic cavity formed by a metallic nanorod integrated into a high-refractive-index dielectric microresonator, we are able to incorporate the highly structured and dispersive character of the electromagnetic spectrum into our analysis, conditions that deviate significantly from the assumptions inherent to conventional spontaneous emission models. Through a detailed comparison with exact solutions that incorporate few quantized field modes, we have benchmarked the effective approach across a range of key parameters, including driving strength and laser frequency detuning. Our results demonstrate that while the effective framework remains remarkably robust in certain degenerate and non-degenerate regimes, its accuracy degrades notably in the presence of strong coherent driving and detuned interactions, where non-Markovian effects and structured photonic baths play a significant role. We believe that our findings extend the understanding of photon-mediated interactions in nanophotonic platforms and highlight both the strengths and limitations of effective open quantum system approaches in non-trivial electromagnetic environments. They also provide practical guidance for their application in quantum technologies involving strongly confined light fields, such as quantum plasmonics and nanocavity QED. 

\begin{acknowledgments}
This work has been funded by MICIU/AEI/10.13039/501100011033 and FEDER, EU under grants PID2021-126964OB-I00 and PID2021-125894NB-I00. AFD and FGV also acknowledge financial support from the European Union’s Horizon Europe Research and Innovation Programme through agreement 101070700 (MIRAQLS). AGT acknowledges support from CSIC Research Platform on Quantum Technologies PTI-001 and from Spanish projects PID2021127968NB-I00 funded by MICIU/AEI/10.13039/501100011033 and by FEDER Una manera de hacer Europa, and from the QUANTERA project MOLAR with reference PCI2024-153449 and funded MICIU/AEI/10.13039/501100011033 and by the European Union.
\end{acknowledgments}

\bibliography{main}

\end{document}